\documentclass[pre,groupedaddress,twocolumn,a4paper,showpacs]{revtex4-1}
\usepackage{graphics}
\usepackage{amssymb}
\usepackage{wasysym}
\usepackage{latexsym}
\usepackage{graphicx}
\usepackage{epsfig}
\usepackage{subfigure}
\usepackage{float}

\begin{document}

\title{Evolving division of labor in a response threshold model}

\author{Jos\'e F.  Fontanari}
\affiliation{Instituto de F\'{\i}sica de S\~ao Carlos, Universidade de S\~ao Paulo,  13560-970 S\~ao Carlos, S\~ao Paulo, Brazil}

\author{Viviane M. de Oliveira}
\affiliation{Departamento de F\'{\i}sica, Universidade Federal Rural de Pernambuco,  52171-900 Recife, Pernambuco, Brazil}

 \author{Paulo R. A.  Campos}
\affiliation{Departamento de F\'{\i}sica, Centro de Ci\^encias Exatas e da Natureza, Universidade Federal de Pernambuco,  50740-560 Recife, Pernambuco, Brazil}

%\date{\today}% It is always \today, today,
             %  but any date may be explicitly specified

\begin{abstract}
 The response threshold model explains the emergence of division of labor (i.e., task specialization) in an unstructured population by assuming that the individuals have  different propensities to work on different  tasks. The incentive to attend to a particular  task increases when the task is left unattended and decreases when individuals work on it. Here we derive  mean-field equations for the stimulus dynamics and show that they exhibit complex attractors  through period-doubling bifurcation cascades when the noise disrupting the thresholds is small. In addition, we  show how the fixed threshold can be set to ensure specialization in  both the transient and  equilibrium regimes of the stimulus dynamics. However, a complete  explanation of the emergence of division of labor requires that we address  the question of where the threshold variation comes from,  starting from a homogeneous population. We then study a structured population scenario, where the population is divided into  a large number of independent groups of equal size, and  the fitness of a group is proportional to the weighted mean work performed  on the tasks during a fixed period of time. Using a winner-take-all strategy to model group competition and assuming an initial homogeneous metapopulation, we find that  a substantial fraction of  workers specialize in each task, without the need to penalize task switching.  
\end{abstract}

%\keywords{Suggested keywords}%Use showkeys class option if keyword
                              %display desired

\maketitle

%
%-----------------------------------------------------
\section{Introduction} \label{sec:intro}
%-----------------------------------------------------
%

Division of labor is a fundamental concept that drives cooperation in human societies \cite{Durkheim(1997)}, colonies of insects \cite{AbbotAnnRevEnt2022}, and bacterial communities \cite{GestelMicBiof2015}. It also plays a central role in the evolution of multicellularity \cite{AmadoEvolution2018, BonnerEvolution2004,TraxlerCurrOpMic2022}, making it an important factor in understanding social organization. Division of labor, which is rooted in promoting efficiency, has been the subject of extensive study and research \cite{AmadoPhysicaA2018, CoopereLife2021, CooperNatEcolEvol2018}.

The prevalence  of the division of labor in social systems warrants   an evolutionary explanation. This explanation must acknowledge that evolution takes place not only among individuals within populations, but also among higher-lever entities, such as  groups or communities, that emerge from their aggregation. Specifically, fitness must be transferred from the individual level to the group level  and eventually return back to the  individuals \cite{MichodPNAS2006}. From this perspective,  the ecological success of specialization at the individual level depends on increasing these returns, which occurs when 
the disruptive costs associated with performing different functions  simultaneously \cite{IspolatovPRSocB2012} and with task switching \cite{,Goldsby(2012),WyliePsycholRes2000} are reduced.  In fact, experiments conducted on eusocial insects demonstrate that the  effectiveness  of individuals  in carrying out specific tasks like foraging or navigation,  increases over time as they become specialized \cite{CollettCurrOpnNeur2003,DukasAnnRevEnt2008,RobinsonAnnRevEnt1992}.

The division of labor among non-reproductive workers is a potential determinant of ecological success in eusocial insects  \cite{GruterPNAS2012}. Eusocial insects are known for their capacity to flexibly respond and adapt to both environmental and internal pressures \cite{GordonAnnRevEnt2019,TheraulazProcRSocB1998}. The behavioral plasticity of individuals determines the flexibility at  the colony level, which in turn affects evolutionary outcomes in changing environments and increases the likelihood of survival \cite{PageAdvInsPhys1991}. In order for the division of labor to be successful at the colony level, workers' actions need to be coordinated, which require some degree of self-organization  \cite{PageApidologie1998,SendovaPhilTrans1999}. Self-organization is attained, for instance,  by  assuming that individuals have an innate threshold for responding to task-related stimuli. Individuals tend to perform a task when the intensity of a stimulus exceeds their innate threshold. Therefore, variation in response thresholds within a population has been suggested as a potential mechanism to explain the emergence of  division of labor in eusocial insects \cite{BeshersAnnRevEnt2001,RobinsonAnnRevEnt1992}.

This conjecture was supported by the response threshold model \cite{Bonabeau(1996),Bonabeau(1997),Bonabeau(1998)}, which assumes a priori that each individual in the colony has a different response threshold for each task: the lower an individual's threshold for a given task, the greater the chance that it will work on that task. In addition, the stimulus intensities of unattended  tasks steadily  increase over time, so that individuals are eventually  prompted to work on the tasks. However, given that variation in  individuals' response thresholds leads to a division of labor, we need to address the question of where the variation comes from, i.e.,  how it can evolve from a homogeneous population. To answer this question, we need to assign each colony a fitness value that summarizes the total amount of work  its members do  on each task, and then allow the colonies to  compete with each other  \cite{Duarte(2012)}. Somewhat surprisingly, this approach failed to produce colonies with a division of labor, i.e., with task specialists. Task specialization seemed to require the introduction of a penalty for task switching, which greatly devalues the claims of the emergence of a division of labor. Since the response threshold model is likely to apply to communities other than eusocial insects,  we will henceforth replace the term colony with group or community.

In this paper,  we re-examine the evolution of the division of labor in a structured population scenario, where the population is divided into  a large number of independent groups of equal size.  Within each group, the stimulus values are determined by the  response threshold model and  the fitness of a group is proportional to the weighted mean work performed  on the tasks during a fixed period of time.    The tasks contribute differently to the fitness. We consider only two tasks, but our results can  be  easily generalized to an arbitrary number of tasks. Starting from a homogeneous metapopulation, we find that when the group dynamics reach the fitness global optimum, which requires a non-genetic elitist optimization strategy \cite{Baluja(1995),Maass(2000)}, a substantial fraction of  workers specialize in each task, in contrast to a previous study that found no such specialization at all \cite{Duarte(2012)}.  

 The remainder of this paper is organized as follows.  In Section \ref{sec:unstr} we revisit the response threshold model in an unstructured population. In particular, we derive  mean-field equations for the stimulus values of the tasks, which  show a complex dynamic behavior when the noise disrupting the thresholds is small.  In this case, bifurcation diagrams show that  the  dynamics produce complex attractors  through period-doubling bifurcation cascades. In addition, by considering fixed heterogeneous thresholds,  we discuss the conditions necessary to ensure specialization in  both the transient and  equilibrium regimes of the stimulus dynamics. In Section \ref{sec:str}, we  consider a structured population scenario to address the question of how the response thresholds can evolve from a homogeneous population to ensure  division of labor with a predetermined distribution of tasks. We show that standard evolutionary algorithms \cite{Goldberg(1989)} fail to reach the global fitness  optimum due to the small  fitness differences between the competing groups. Finally, in Section \ref{sec:disc} we discuss what biases the winner-take-all algorithm, which we  used in the group competition dynamics,  toward certain optima of the fitness function.

%
%-----------------------------------------------------
\section{Unstructured population}\label{sec:unstr} 
%-----------------------------------------------------
%

The unstructured population  dynamics is governed by the classical response threshold model \cite{Bonabeau(1996),Bonabeau(1997),Bonabeau(1998)}. Each agent $i=1, \ldots, N$  is characterized by  the  thresholds  $\theta_i^a $ for tasks $a=1,2$, and each task is characterized by a time-dependent stimulus $S_a (t) $.  Agent $i$  works on task  $a$ at time step  $t=0, \ldots,T-1$ if  the inequality
\begin{equation}\label{ineq}
\theta_i^{a} < S_a (t)  + \epsilon_i^a (t)
\end{equation}
is verified, where $\epsilon_i^a (t)$ is a normally distributed random variable with zero mean and variance $\sigma^2$. In the case that the inequality (\ref{ineq}) is  verified  for both tasks $a =1$ and $a =2$,  agent $i$ chooses one of the tasks at random and works on it. Thus, an agent can deal with at most one task  per time step. Furthermore, agent $i$ is idle at time step $t$ if the inequality (\ref{ineq}) is violated for both tasks. More precisely, introducing the sigmoid function
\begin{equation}\label{Phi}
\Phi_i^a(t) = \frac{1}{2} \left [ 1 + \mbox{erf} \left ( \frac{S_a(t) - \theta_i^a}{\sqrt{2 \sigma^2}} \right ) \right ],
\end{equation}
where $ \mbox{erf}(x)$ is the error function, we write  the probability that agent $i$  works on task 1 at time step $t$ as 
$\Phi_i^1 (1- \Phi_i^2) + \Phi_i^1 \Phi_i^2/2$, that it works on  task 2 as $\Phi_i^2 (1- \Phi_i^1) + \Phi_i^1 \Phi_i^2/2$ and that it is idle as $(1-\Phi_i^1) (1- \Phi_i^2) $. 

The interesting feature of the  response threshold model is that the stimulus $S_a$  changes over time depending on the amount of work done on task $a$. In particular,  $S_a$ decreases of the fixed amount $\alpha_a$ for each agent working on  task $a$  at time step $t$. On the other hand, $S_a$ increases of the fixed amount  $\delta_a$, regardless of whether  task $a$ is  attended or not. Since the aim here is to study the emergence of specialization, we do  not  want to distinguish the tasks a priori, so we set $\delta_a = \delta$ and $\alpha_a = \alpha/N$ for $a=1,2$. The scaling factor $N$ is introduced so as to make the results less sensitive to changes in the group size $N$. The stimulus dynamics is thus described by the equation 
\begin{equation}\label{stim}
S_a (t+1) = S_a (t) + \delta - \alpha \eta_a (t) ,
\end{equation}
where 
\begin{equation}
\eta_a (t)  = \frac{ A_a (t)}{N}
\end{equation}
  and $A_a(t)$ is the number of agents that work on task $a$ at time step $t$. Since $A_a(t)$ scales with $N$, our rescaling of  $\alpha_a$ guarantees that $\alpha$ is a parameter of order of one. In fact,  we will set $\alpha=1$ throughout the paper without loss of generality. This can be done by measuring $S_a$, $\theta_i^a$, $\delta$ and $\sigma$ in units of $\alpha$. We note that in this setting, the parameters used in the study by Duarte et al. are $N=100$, $\delta =  1/3$ and $\sigma^2 = 1/9$ \cite{Duarte(2012)}.

The equilibrium analysis of eq.\  (\ref{stim}) is very instructive. In fact, setting $S_a (t+1) = S_a (t)$  gives $\eta_a =  \delta$  for $a=1,2$. Thus, at equilibrium the number of agents working on  each task is equal, which is expected since our choice of parameters (i.e., $\delta_a = \delta$ and $\alpha_a =  \alpha/N$)  makes the two  tasks indistinguishable. As a result, the proportion of inactive agents  at equilibrium is $1 - 2\delta$, and  consequently eq.\  (\ref{stim}) has no  equilibrium solution  for  $\delta > 1/2$. What happens in this case?  Both stimuli diverge as $(\delta-1/2) t$ for $t$ large so that the agents are always ready to work on both tasks. Thus, each task is attended on average $N/2$ times  at each time step. Furthermore,  the agents switch tasks with probability $1/2$ at each time step.  However, the situation for 
 $\delta \leq  1/2$ is very different, as we will see next. In particular,  although the condition $\eta_1=\eta_2 = \delta$ is satisfied at equilibrium, some agents may only work on a single task and some may remain idle forever: it all depends on the (fixed) values of the thresholds $\theta_i^a$.
 
The stochastic simulations of the unstructured population dynamics are implemented as follows.  Throughout the paper, we set the initial stimulus values to $S_a (0) = 10$ for $a=1,2$. This choice is inconsequential in our setting, as we do not  impose any arbitrary constraints on the stimulus values  resulting from eq. (\ref{stim}), unlike to  Duarte et al.  that requires the stimulus values to be nonnegative \cite{Duarte(2012)}. Given the fixed thresholds  $\theta_i^a$, we  check the inequalities (\ref{ineq}) for each agent in the population. In doing so, we determine the proportion of agents that work on each task (i.e., $\eta_a (0)$ for $a=1,2$) so that we can evaluate the stimulus values at time step $t=1$ using eq.\  (\ref{stim}). Once $S_a(1)$ for $a=1,2$  are known, we can check the inequalities (\ref{ineq})  again and calculate $\eta_a(1)$ for $a=1,2$, which allows us to obtain  $S_a(2)$ for $a=1,2$. This procedure is then repeated for an arbitrary number of time steps.

Note that for large $N$ we can easily write a set of deterministic recurrence equations for $S_a(t)$ using the approximations
\begin{eqnarray} 
 \eta_1 (t)  & \approx  &  \frac{1}{N} \sum_{i=1}^N \left [ \Phi_i^1(t)  \left [ 1- \Phi_i^2(t) \right ] + \frac{1}{2} \Phi_i^1(t) \Phi_i^2(t) \right ] \nonumber \\ 
 \eta_2 (t) & \approx  &  \frac{1}{N} \sum_{i=1}^N \left [ \Phi_i^2 (t) \left [ 1- \Phi_i^1(t) \right ] + \frac{1}{2} \Phi_i^1 (t) \Phi_i^2(t) \right ]
 \end{eqnarray}
in eq. (\ref{stim}), where $\Phi_i^a(t)$ is given by eq. (\ref{Phi}).  Next, we will compare the predictions of this mean-field-like  approximation with the results of the stochastic simulations for agent-independent (homogeneous) thresholds and for  agent-dependent (heterogeneous) thresholds.

\subsection{Homogeneous thresholds}\label{sec:homo} 

\begin{figure}[t] 
\centering
 \includegraphics[width=.8\columnwidth]{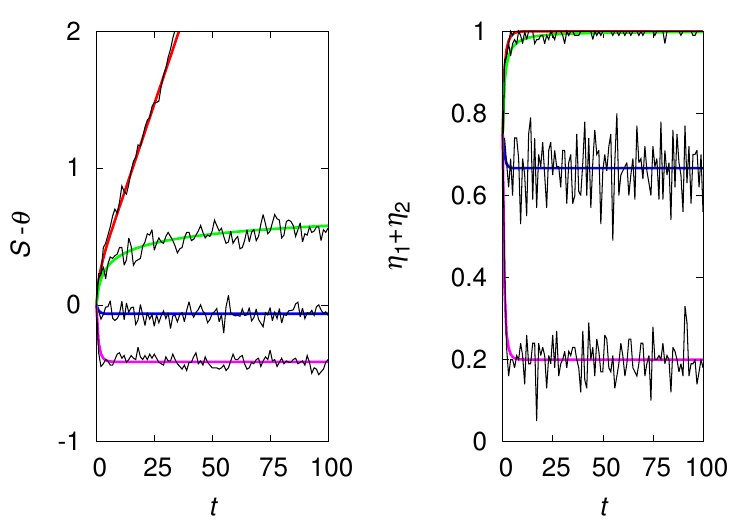}  
\caption{Time evolution of the shifted stimulus values $S-\theta$  (left panel) and  proportion of active agents $\eta_1 + \eta_2$ (right panel)  
for (from top to bottom) $\delta=0.55, 0.5, 1/3$ and $0.1$ in the case of homogeneous thresholds.  The colored thick lines are the results of the deterministic approximation  and the black thin lines are the results of a single stochastic simulation.
The other parameters are $N=100 $ and  $\sigma^2=1/9$ and the initial condition is $S(0) = \theta$.
 }  
\label{fig:1}  
\end{figure}

In this case we set $\theta_i^a = \theta$ for $a=1,2$ and $i=1, \dots,N$.   Since the initial condition is $S_1(0) = S_2(0) = 10$, there is nothing to distinguish between the different agents and tasks, so we can set $S_1(t) =  S_2(t) = S(t) $  for all time steps.  Fig.\ \ref{fig:1} shows the results of  iterating  the recurrence equation
\begin{equation}\label{Sh}
S (t+1) = S (t) + \delta - \frac{1}{2} + \frac{1}{2} \left [ 1 - \ \Phi(S) \right ]^2
\end{equation}
with 
\begin{equation}\label{Phi}
\Phi (S) = \frac{1}{2} \left [ 1 + \mbox{erf} \left ( \frac{S (t) - \theta}{\sqrt{2 \sigma^2}} \right ) \right ],
\end{equation}
and of a single agent-based stochastic simulation. In the simulations the agents and tasks keep their identity by construction. Note that by subtracting $\theta$ on both sides  of eq.\  (\ref{Sh}), we obtain an equation for the shifted variable  $S-\theta$, so that the choice of $\theta$ becomes inconsequential for our analysis. The results confirm the linear divergence of the stimulus with  $t$ for $\delta > 1/2$.  For  $t=0$ we have $\eta_1=\eta_2 =  3/8$ so the  proportion of active agents is $\eta_1+ \eta_2 =  3/4$. For $t \to \infty$,  this  proportion is $\eta_1+ \eta_2 = 2 \delta$ for $\delta \leq 1/2$ and $\eta_1+ \eta_2 = 1$ for $\delta > 1/2$, as expected. The shifted stimulus values at equilibrium $S_{eq} - \theta$ can be determined by the condition  
\begin{equation}\label{Phi_eq}
\Phi_{eq} =  1 - \left ( 1 - 2\delta \right )^{1/2},
\end{equation}
from which we obtain $S_{eq} -  \theta= \sqrt{2\sigma^2} \mbox{erf}^{-1} (2\Phi_{eq} -1) $.  Note that $S_{eq}$ diverges  for $\delta=1/2$.  It is interesting that for, e.g.,  $\delta =1/3$ we have $S_{eq} -\theta  \approx - 0.07 $, so  it is the noise $ \epsilon_i^a $ in eq.\ (\ref{ineq}) that allows the agents to work on the tasks. The value $S_{eq}$ is such that each agent attends  each task with  probability $\delta$ and is idle with probability $1- 2 \delta$. Thus all agents are generalists in the case of homogeneous thresholds.

%-----------------------------------------------------
%-----------------------------------------------------

Most interestingly, although the  equilibrium solution discussed above exists for all values of  $\sigma^2$ and $\delta < 1/2$, it is unstable for small noise variances. In fact, by defining $F(S) = S  + \delta -1/2 +  \left ( 1 -  \Phi \right )^2/2$ the condition for the local stability of the equilibrium solution $S = S_{eq}$ is \cite{Strogatz(2014)}
\begin{equation}
\left  | \frac{dF}{dS}\mid_{S_{eq}} \right  |  < 1 
\end{equation}
where
\begin{equation}
 \frac{dF}{dS}\mid_{S_{eq}}  = 1 - \left ( \frac{1 - 2\delta}{2 \pi \sigma^2} \right )^{1/2} \exp  \left [ -\frac{(S_{eq} - \theta)^2}{2 \sigma^2} \right ] .
\end{equation}

\begin{figure}[t] 
\centering
 \includegraphics[width=.8\columnwidth]{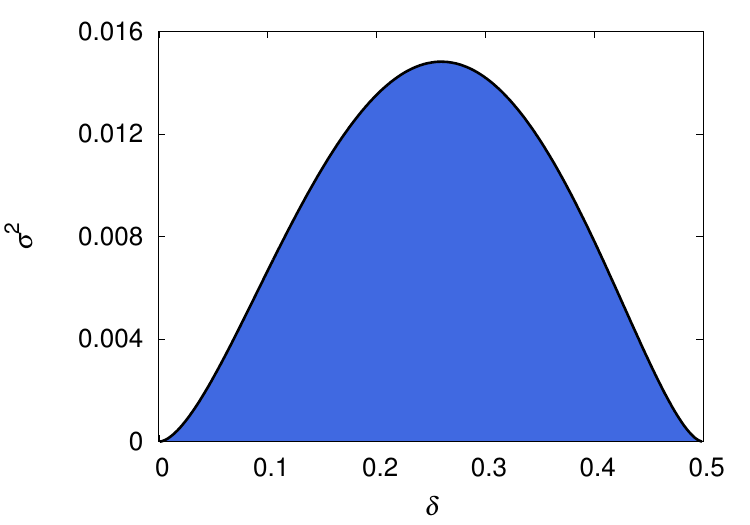}  
\caption{Phase diagram in the parameter space $(\delta,\sigma^2)$ showing the region of instability (shaded region) of the fixed-point solution $S=S_{eq}$ in the case of homogeneous thresholds.  In the shaded region the attractors of the dynamics are period-$n$ cycles with $n=2,3, \ldots$ and aperiodic cycles.
 }  
\label{fig:2}  
\end{figure}

Fig.\ \ref{fig:2} shows the region of instability of the fixed-point solution $S=S_{eq}$ of eq.\ (\ref{Sh}).  What happens in this region? Probably the best  way to understand the outcome of the dynamics in this region is  to draw bifurcation diagrams \cite{Strogatz(2014)}:  starting from the same initial condition $S(0)=\theta$, we iterate eq. (\ref{Sh}) for long enough to guarantee that the dynamics enters the stationary regime  and then we plot the values of $S(t) - \theta$ for 100 consecutive time steps. In this way we can visualize the  attractors of the dynamics as a function of the model parameters $\delta$ and $\sigma^2$.

%-----------------------------------------------------
%-----------------------------------------------------

\begin{figure}[t] 
\centering
 \includegraphics[width=.8\columnwidth]{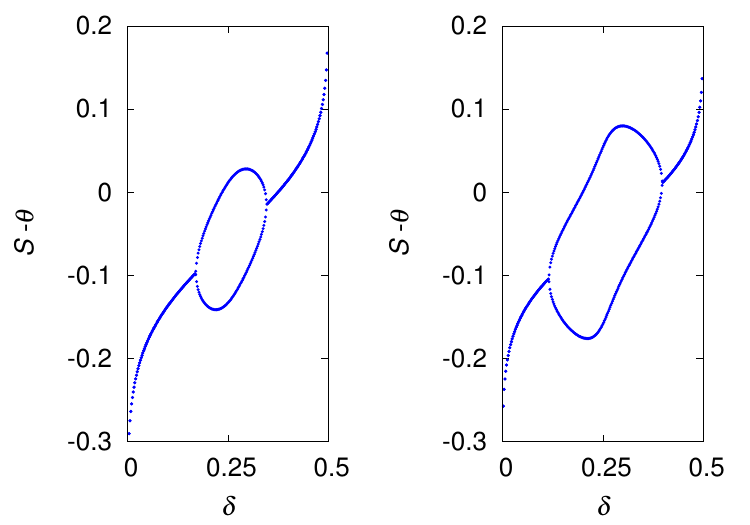}  
\caption{Bifurcation  diagram of the  variable $S-\theta$  for $\sigma^2 = 0.012 $ (left panel) and $\sigma^2 = 0.008 $ (right panel). For a not too small noise variance, the attractors are fixed points or period-$2$ cycles.
 }  
\label{fig:3}  
\end{figure}

\begin{figure}[h!] 
\centering
 \includegraphics[width=.8\columnwidth]{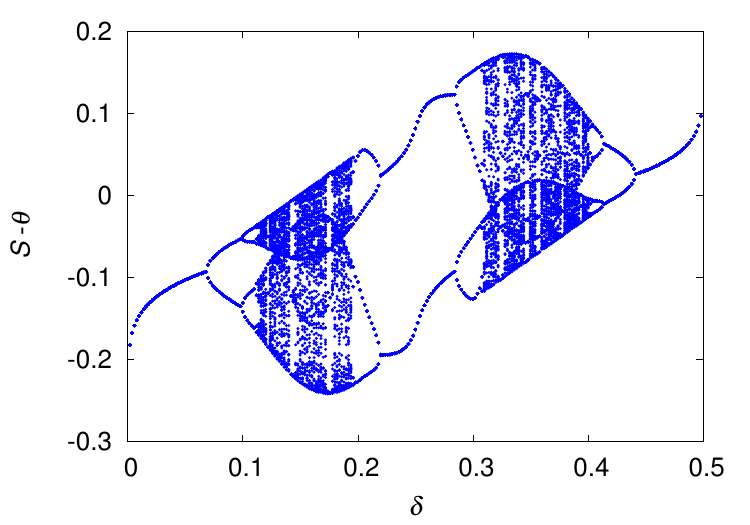}  
\caption{Bifurcation diagram of  the  variable $S-\theta$  for $\sigma^2 = 0.004 $. For  small noise variance, complex attractors appear through period-doubling bifurcation cascades.
 }  
\label{fig:4}  
\end{figure}

Fig.\ \ref{fig:3} shows the bifurcation diagrams for high values of the noise variance  (but low enough to destabilize the fixed-point solutions). In this case, the  attractors in the shaded region of  Fig.\ \ref{fig:2} are only  period-$2$ cycles. The maximum amplitude of the cycle occurs for $\delta = 1/4$. However, decreasing  $\sigma^2$ leads to the appearance of  much  more complex attractors  through period-doubling bifurcation cascades, as shown in Fig.\ \ref{fig:4}.  It is interesting to note  that for $\delta=1/4$ the attractors are fixed points or period-$2$ cycles, regardless of the noise variance $\sigma^2$. The period-$2$ cycles for small $\sigma^2$ are easy to understand in this case because when a task is attended,  its shifted stimulus is decremented  of $1/2 -1/4 = 1/4$, whereas when this  task is not attended, it is incremented of $1/4$, so  after two time steps the shifted stimulus returns to its initial value resulting in the observed period-$2$ cycle.

 %-----------------------------------------------------
%-----------------------------------------------------
 
 %-----------------------------------------------------
%-----------------------------------------------------

\begin{figure}[t] 
\centering
 \includegraphics[width=.8\columnwidth]{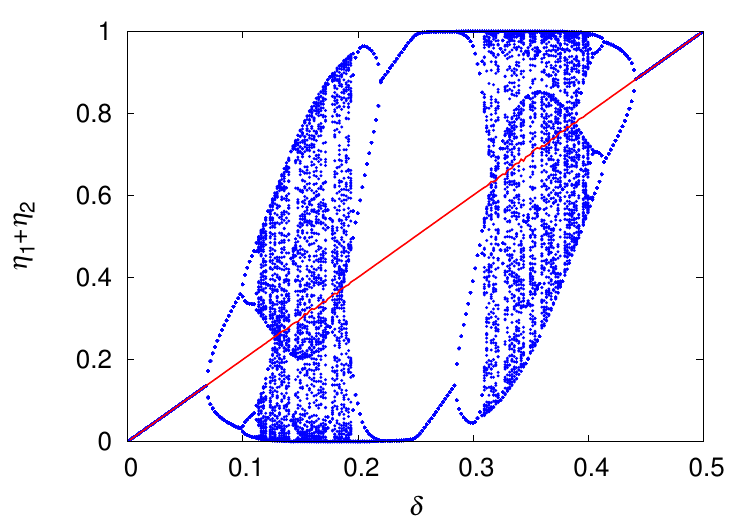}  
\caption{Bifurcation diagram of the   proportion of active agents $ \eta_1 + \eta_2 = 1 - (1-\Phi)^2 $  for $\sigma^2 = 0.004 $. The solid line is the average over the points in the attractor, which coincides with the result for the fixed-point attractor, viz. $\eta_1 + \eta_2 = 2 \delta$.
 }  
\label{fig:5}  
\end{figure}

An interesting question is whether the complex attractors of the recurrence (\ref{Sh}) can actually increase the amount of work done by the agents as compared to the fixed point solution. We address this point by looking at the  proportion of active agents $ \eta_1 + \eta_2 = 1 - (1-\Phi)^2 $ for the attractors shown in the bifurcation diagram of Fig.\ \ref{fig:4}, which is shown in Fig.\ \ref{fig:5} together with the average activity for each attractor. Remarkably, if  we average the  proportion of active agents over all values visited by the dynamics in the stationary regime, we obtain  the same value of the fixed-point solution, i.e., $\eta_1 + \eta_2 = 2 \delta$. So the total  work done at equilibrium  is $2 \delta$, regardless  of the values of $\sigma^2$ and $\delta$.

%-----------------------------------------------------
%-----------------------------------------------------

\subsection{Heterogeneous thresholds}\label{sec:heter}

In the seminal  paper on the response threshold model, it was  assumed that there were two classes of agents (in this case, ant castes)  characterized by different thresholds for the existing tasks \cite{Bonabeau(1996)}. This approach has been criticized in the context of the emergence of the division of labor, since specialization is assumed a priori \cite{Duarte(2012)}. We believe that  this criticism is unjustified, since the  division of labor ultimately depends  on the outcome of the stimulus dynamics  (\ref{stim}), and  it is therefore  very difficult to know a priori whether a  particular  assignment  of thresholds to  the tasks  will  lead to specialization. Furthermore, the study of the response threshold model with heterogeneous thresholds  provides invaluable clues to understanding the evolution of specialization in a group selection scenario.

Consider a threshold assignment  such that a  proportion $\beta$ of  $N$ agents  works on task $1$ (class 1),  a  proportion  $1-\beta$ works on task $2$ (class 2) and  no agent is idle during  a given number $T$ of time steps.   We assume here, without loss of generality, that $\beta \leq 1/2$: the case $\beta > 1/2$ is obtained by simply interchanging the  labels of the  tasks.  Assuming  $\beta < \delta$ and  $\sigma^2 =0$, it is easy to check  that the assignment 
\begin{eqnarray}\label{t_a}
\theta_i^1 &  < &  S_1(0) \nonumber \\
\theta_i^2 &  >  & S_2(0) 
\end{eqnarray}
for $i=1, \ldots, \lfloor \beta N \rfloor $ and
\begin{eqnarray}\label{t_b}
\theta_i^1 &  > &  S_1(0) + T (\delta - \beta)  \nonumber \\
\theta_i^2 &  <  & S_2(0)  -  T (1 - \beta - \delta)
\end{eqnarray}
for $i=\lfloor \beta N \rfloor + 1, \ldots,  N$ guarantees this division of labor for the first $T$ time steps. Here  $ \lfloor x \rfloor$ stands for greatest integer less than or equal to $x$.  For $\beta > \delta$ and  $\sigma^2 =0$, the corresponding assignment is
\begin{eqnarray}\label{t_c}
\theta_i^1 &  < &  S_1(0)  - T (\beta - \delta) \nonumber \\
\theta_i^2 &  >  & S_2(0)  
\end{eqnarray}
for $i=1, \ldots, \lfloor \beta N \rfloor$ and
\begin{eqnarray}\label{t_d}
\theta_i^1 &  > &  S_1(0)   \nonumber \\
\theta_i^2 &  <  & S_2(0)  -  T (1 - \beta - \delta)
\end{eqnarray}
for $i=\lfloor \beta N \rfloor  + 1, \ldots,  N$. In writing these inequalities we  have used that $1 -\beta - \delta \geq 0$  since $\delta \leq 1/2$  and $\beta \leq 1/2$. Of course, the choice of which agents belong to each of the two classes is immaterial, but our choice greatly simplifies the notation.

\begin{figure}[t] 
\centering
 \includegraphics[width=1\columnwidth]{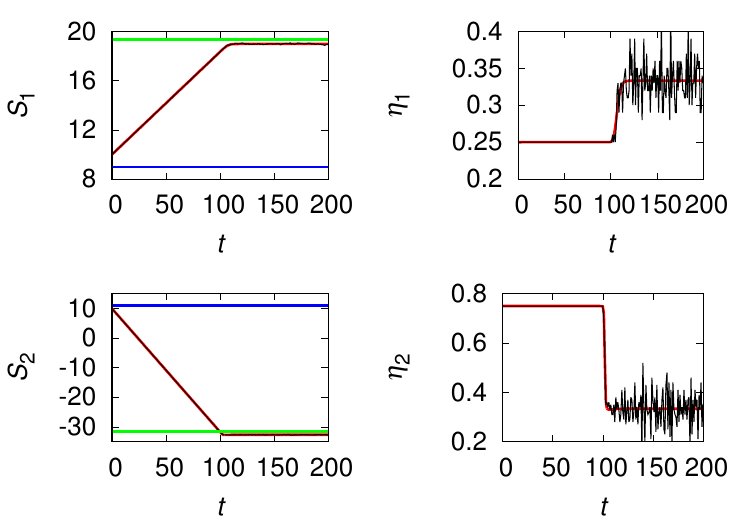}  
\caption{Time evolution of the  stimulus values of task 1 (upper left panel) and  task 2 (lower left panel), and of the   proportion of  agents that work on task 1  (upper right panel)  and task 2 (lower right panel). The horizontal blue lines are the thresholds $\theta_i^1 = 9$ and $\theta_i^2 = 11$ for $i=1, \ldots, 25$, whereas the horizontal green lines are the thresholds $\theta_i^1 = 19.33$ and $\theta_i^2 = -31.66$ for $i=26, \ldots, 100$.
The red curves are the results of the mean-field approximation  and the black thin lines are the results of a single stochastic simulation.
The parameters are $N=100 $, $T=100$, $\beta =1/4$, $\delta= 1/3$  and  $\sigma^2=1/9$. The initial stimulus values are $S_1(0) = S_2(0) =10$.
 }  
\label{fig:6}  
\end{figure}

Fig.\ \ref{fig:6} summarizes our results for $\beta = 1/4 < \delta = 1/3$ and $T=100$. It is clear that even in the presence of noise (i.e., $\sigma^2 =1/9$), the prescription (\ref{t_a}) and (\ref{t_b}) guarantees a perfect  division of labor in the transient regime ($0 < t < T-1$). In this regime, a  proportion $\beta$ of agents are tuned only to task 1,  while the reminder is tuned only to task 2. In the equilibrium regime, agents in class 1 continue to work exclusively on  task 1, but agents in class 2 now work on task 1 with probability $\delta - \beta$, on task 2 with probability $\delta$ and remain idle with probability $1- 2 \delta$.  This is necessary to satisfy the equilibrium condition 
$\eta_1 = \eta_2 = \delta$.  The results of a single run of the stochastic simulation algorithm are also shown in Fig.\ \ref{fig:6}: they are  indistinguishable from the  mean-field approximation in the transient regime. In addition, the large values of the stimulus values make the stochastic and the mean-field results indistinguishable on  the scale of the figures in the equilibrium regime as well. In summary, for the setting of parameters of Fig.\ \ref{fig:6},  our prescription for the thresholds produced a group of specialists on task 1 but not on task 2 in the equilibrium regime. This result illustrates our point that it is difficult to set the  thresholds   to ensure a perfect division of labor in equilibrium. However, we can create a perfect division of labor in the case $\beta = 1/2$, as we show next.

%-----------------------------------------------------
%-----------------------------------------------------

\begin{figure}[t] 
\centering
 \includegraphics[width=1\columnwidth]{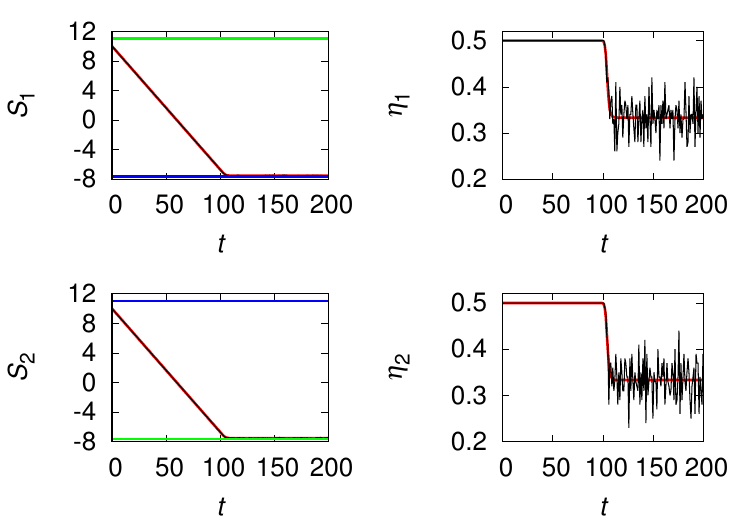}  
\caption{Time evolution of the  stimulus values of task 1 (upper left panel) and  task 2 (lower left panel), and of the  proportion of  agents that work on task 1  (upper right panel)  and task 2 (lower right panel). The horizontal blue lines are the thresholds $\theta_i^1 = -7.66$ and $\theta_i^2 = 11$ for $i=1, \ldots, 50$, whereas the horizontal green lines are the thresholds $\theta_i^1 = 11$ and $\theta_i^2 = -7.66$ for $i=51, \ldots, 100$.
The red curves  are the results of the mean-field approximation  and the black thin lines are the results of a single stochastic simulation.
The parameters are $N=100 $, $T=100$, $\beta =1/2$, $\delta= 1/3$  and  $\sigma^2=1/9$. The initial stimulus values are $S_1(0) = S_2(0) =10$.
 }  
\label{fig:7}  
\end{figure}

Fig\ \ref{fig:7} summarizes our results for $\beta = 1/2 > \delta = 1/3$ and $T=100$, using the prescription  (\ref{t_c}) and (\ref{t_d}) for the thresholds of the $N=100$ agents. Since in this case there is a symmetry between the two tasks, we find $S_1=S_2$ and
$\eta_1 = \eta_2$ for all time steps $t$. In the transient regime ($0 < t < T-1$), all  agents are active and work on their specific tasks, as expected. Comparing the thresholds (horizontal lines in the left column of Fig\ \ref{fig:7}) with the stimulus values makes it clear that  agents  in class 1 (i.e., agents $i=1,\ldots, 50$) will never work on task 2 and agents in class 2 (i.e., agents $i=51,\ldots, 100$) will never work on   task 1. At equilibrium we therefore have a perfect division of labor,  where an agent always works on the same task when it is active. Thus, in this scenario all agents are specialists in their tasks.  Since in equilibrium we have $\eta_1 = \eta_2$, we have the impression  that this fortunate outcome is only  possible for $\beta=1/2$ when the tasks are equivalent at the outset.   

%-----------------------------------------------------
%-----------------------------------------------------

%
%-----------------------------------------------------
\section{Structured population}\label{sec:str} 
%-----------------------------------------------------
%

In the previous section we showed   how the division of labor can emerge if the thresholds for the different tasks are properly chosen. Now we address the important question of how these thresholds can evolve from a homogeneous population to  give rise to a division of labor with a preset distribution of work across tasks. A natural  and sensible approach, suggested by Duarte et al.,  is to  
divide the population into $M$ independent groups of $N$ individuals each, and allow the groups to compete with each other \cite{Duarte(2012)} (see also \cite{Oliveira(2019)}).  Within each group, the stimulus values are determined by the  response threshold model.  The fitness $w(T$) of a group is given  by the properly weighted and scaled mean work performed  on the two tasks during  $T$ time steps, i.e.,  
\begin{equation}\label{w}
w(T) = \left [   \bar{\eta}_1 (T)  \right ]^\beta  \left [   \bar{\eta}_2 (T)  \right ]^{1-\beta},
\end{equation}
where
\begin{equation}\label{bareta}
  \bar{\eta}_a (T) =  \frac{1}{T} \sum_{t=0}^{T-1} \eta_a (t) 
\end{equation}
for $a=1,2$ and 
the  exponent $0 < \beta \leq 1/2 $ is a weighting factor indicating the relative importance of tasks 1 and 2 \cite{Duarte(2012)}. For example,  for $\beta = 1/2$ both tasks are equally important, whereas for $\beta = 1/4$ the optimal strategy for the group requires that task 2 be  performed three times more often than task 1.  We note that the generalized geometric mean  appearing in eq. (\ref{w}) is the usual choice to enforce the coexistence of replicators in  group selection models of prebiotic evolution, where each replicator  type specializes on the production of a  metabolite essential for the survival of the protocell \cite{Alves(2000),Czaran(2000),Fontanari(2006)}.  This is a scenario of dependent specialization, in which the survival of the specialists depends on the community \cite{Wilson(1980)}.

Fitness $w(T)$ is maximized if there are no idle agents  during the $T$ time steps when the total work  is measured. In this case, we have $\eta_1(t) + \eta_2(t) = 1$ for $t=0, \ldots,T-1$, so that  $\bar{\eta}_2 (T) = 1 -  \bar{\eta}_1 (T)$.  Therefore, maximizing the  fitness  $w(T)$  with respect to $\bar{\eta}_1 (T)$  gives
\begin{equation}\label{wopt}
w_{opt} = \beta^\beta  \left (1-\beta  \right )^{1-\beta}
\end{equation}
for $\bar{\eta}_1 (T) = \beta$. 
It is then clear that our prescriptions for the thresholds (\ref{t_a})-(\ref{t_b}) for $\beta < \delta$ and
(\ref{t_c})-(\ref{t_d})  for $\beta > \delta$ (see Figs.\ \ref{fig:6} and \ref{fig:7})  give a global maximum of the fitness $w(T)$. Hence the use of the same notation $\beta$ for the proportion of agents attending to task 1 in section \ref{sec:heter} and the relative weight of task 1 in eq.\ (\ref{w}). We note, however,  that the global maximum $w_{opt}$ is not unique: There are other ways of  setting $ \bar{\eta}_{1}(T) = \beta$ other than assuming that a fixed  proportion $\beta$ of agents specialize in task 1, as we do  in our  threshold prescriptions.

Interestingly, variants of the standard genetic algorithm \cite{Goldberg(1989)}, where each group contributes with offspring to the next generation  with a probability proportional to its fitness $w(T)$,  failed to produce the expected division of labor \cite{Duarte(2012)}.  Somewhat artificial corrections, such as a penalty for task switching \cite{Goldsby(2012)}, were  considered necessary for the response threshold model to ensure  that the agents stick to a particular task. The reason for this failure is twofold. First, Duarte et al. has arbitrarily constrained the stimulus values and the thresholds to take only  nonnegative values \cite{Duarte(2012)}. However,  it is clear from Figs.\ \ref{fig:6} and  \ref{fig:7} that both the stimulus dynamics (\ref{stim}) and the optimal threshold prescription require that these variables also take on  negative  values. Second and more importantly, the standard genetic algorithm fails to find the global optimum of the fitness (\ref{w}), as discussed next.

By standard genetic algorithm we mean the procedure  by which  a given group, say,  group $k$,  is selected to  reproduce  (asexually) with the probability  $w_k(T)/\sum_{l=1}^M w_l(T)$, where we have  added the subscript $k=1, \ldots,M$ to the fitness (\ref{w}) in order to identify the groups. A total of $M$ groups are simultaneously selected with replacement using this selection procedure. Once a group is selected for reproduction, the thresholds of all agents in the group are slightly modified (i.e., mutated)  by  adding   a random Gaussian noise with mean  $0$ and variance  $0.1$.  Each generation $\tau$ of the group dynamics comprises $T$ time steps of the stimulus dynamics within each group, which is  necessary to calculate the group fitness. In each group generation, the stimulus values are reset to the values $S_1(0)=S_2(0) = 10$ in all groups.  In  the initial generation ($\tau=0$), we set $\theta_i^a = 10$ for all agents and tasks within each group, so that the  $2NM$  thresholds are identical for all agents  at the start of the group dynamics. Thus, the metapopulation is homogeneous at $\tau=0$.

\begin{figure}[t] 
\centering
 \includegraphics[width=.8\columnwidth]{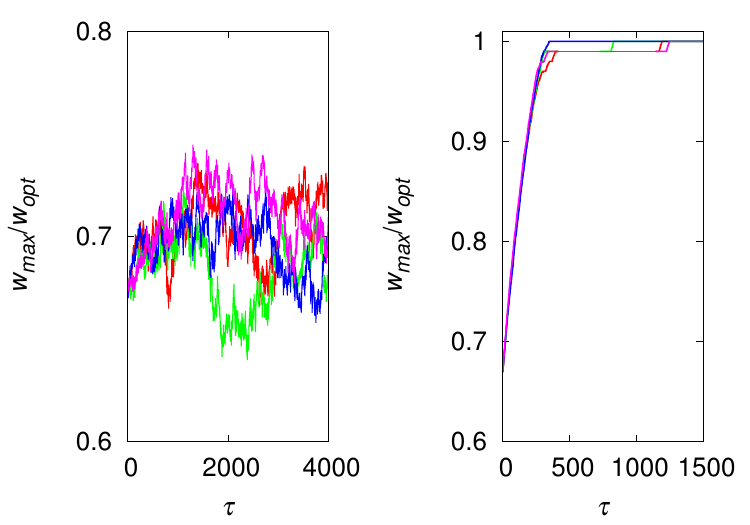}  
\caption{Ratio between the maximum fitness and the theoretical optimal fitness for four runs of the standard genetic algorithm  (left panel) and the winner-take-all  algorithm (right panel) simulating  competition between  $M=1000$ group of $N=100 $ agents each.
The other parameters are $T=100$, $\beta=1/2$ ,$\delta =1/3$ and  $\sigma^2=1/9$.
 }  
\label{fig:8}  
\end{figure}

Fig.\ \ref{fig:8} (left panel) shows the ratio between the maximum fitness value $w_{max}$ at group generation $\tau$ and the optimal fitness $w_{opt}$ for four runs of the standard genetic algorithm described above. The results show that there is almost no improvement over the  initial setting of the metapopulation.  We have verified that this conclusion holds regardless of the number of groups, group generations and mutation procedure. The reason is that the  difference between the maximum and the average fitness is too small to guarantee that the fittest groups will pass their offspring to the next generation, so the offspring are essentially random samples of the previous generation. This situation can be dealt with by introducing an elitist reproduction strategy in which the fittest  group is always guaranteed to contribute a clone to the next generation \cite{Baluja(1995)}. Here we take a more radical stance and assume that the fittest group  produces  all $M$ offspring  and that all thresholds of the agents in these  related groups undergo mutations.  We refer to this reproduction strategy as the winner-take-all  algorithm \cite{Maass(2000)}. Fig.\ \ref{fig:8} (right panel) shows that this algorithm quickly reaches  the optimal fitness value $w_{opt}$. 

%-----------------------------------------------------
%-----------------------------------------------------

\begin{figure}[t] 
\centering
 \includegraphics[width=.8\columnwidth]{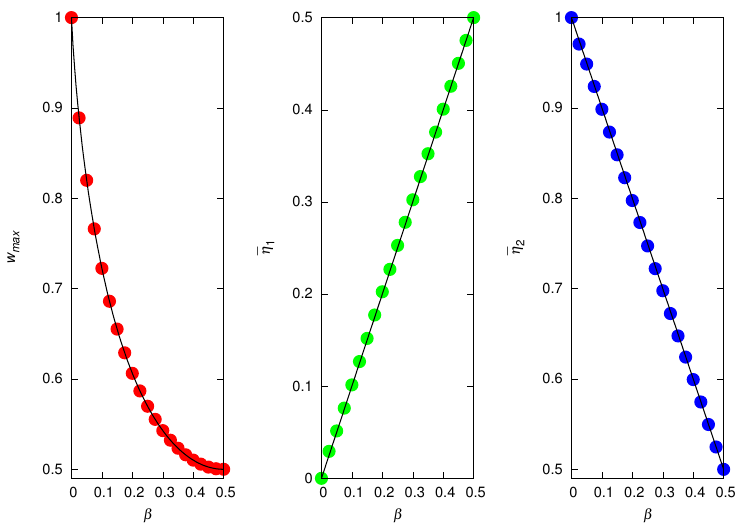}  
\caption{Maximum fitness of the winning  group (left panel),  mean work done on task 1 (middle panel) and  mean work done on task 2  (right panel)  as a function of  the relative importance  $\beta$  of tasks 1 and  2.  The symbols are the averages over 500 runs of the winner-take-all algorithm and the curves are the optimal results, i.e., $w_{max}=w_{opt}$, $\bar{\eta}_{1}= \beta$ and $\bar{\eta}_2 = 1 - \beta$. The parameters are $M=100$,  $N=100 $, $T=100$, $\delta =1/3$ and  $\sigma^2=1/9$.
 }  
\label{fig:9}  
\end{figure}

The question we now address  is whether the winner-take-all  algorithm finds the global fitness maximum  $w_{opt}$ by exploiting the  division of labor or  by using  some other strategy. Fig.\ \ref{fig:9} shows the performance of this algorithm averaged over 500 runs. We let the algorithm run for $10^5$ group generations or until $w_{opt} - w_{max} < 10^{-6}$.  There are two points  to emphasize about these results.  First, since  $\bar{\eta}_{1} +  \bar{\eta}_{2} = 1$,  the maximum of  $w(T)$  is reached when all agents are active during the whole observation period $T$, as expected. Second, maximizing $w(T)$  with $\beta < 1/2$ is sufficient to produce an asymmetric distribution of work across tasks (i.e., $\bar{\eta}_{1} \neq \bar{\eta}_{2}$), in contrast with the results of Duarte et al. that require a clumsy modification of the fitness $w(T)$  to enforce the preset  distribution of work across tasks \cite{Duarte(2012)}.

%-----------------------------------------------------
%-----------------------------------------------------

\begin{figure}[t] 
\centering
 \includegraphics[width=.8\columnwidth]{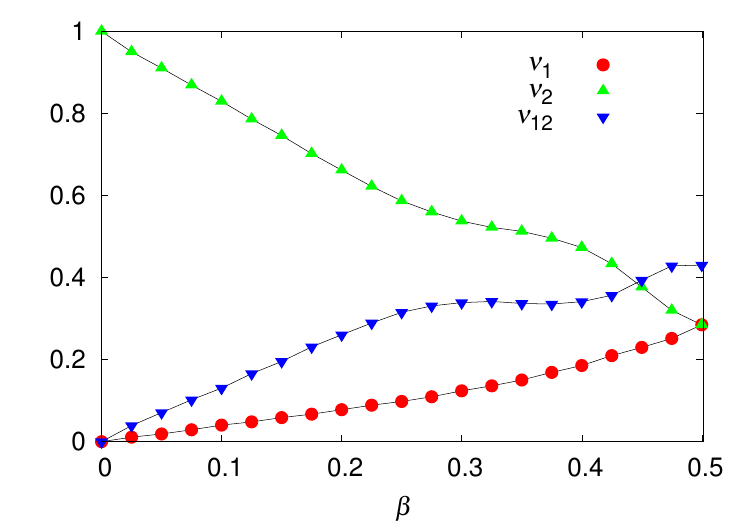}  
\caption{Proportion of agents in the winning group that work exclusively on task 1 ($\nu_1$),  that work exclusively on task 2 ($\nu_2$) and that work on  both tasks ($\nu_{12}$) as  a function of  the relative importance  $\beta$  of tasks 1 and  2. The symbols are the averages over 500 runs of the winner-take-all algorithm and the lines are guides to the eye. The parameters are $M=100$,  $N=100 $, $T=100$, $\delta =1/3$ and  $\sigma^2=1/9$.
 }  
\label{fig:10}  
\end{figure}

Now, to find out whether the winner-take-all algorithm explores the division of labor to optimize the fitness $w(T)$, we measure the  proportion of agents in the winning group that work exclusively on task 1,  denoted by  $\nu_1$,  that work exclusively on task 2, denoted by $\nu_2$, and that work on both tasks, denoted by $\nu_{12}$. Since there are no idle agents during the observation period $T$, we have $\nu_1 + \nu_2 + \nu_{12} = 1$ and the  proportion of specialists is $\chi \equiv \nu_{1}+\nu_2$. 
Fig.\ \ref{fig:10} shows that the  worst case for the  division of labor is $\beta = 1/2$ and even in this case  57\% of the agents work exclusively on one of the tasks. Interestingly, we find  that the generalists attend to both tasks with approximately the same frequency on the average, regardless of the value of $\beta$, so that $\nu_1 + \nu_{12}/2 = \beta$ and $\nu_2 + \nu_{12}/2 = 1 - \beta$.

%-----------------------------------------------------
%-----------------------------------------------------

%-----------------------------------------------------
\begin{figure}[h!] 
\centering
 \includegraphics[width=.8\columnwidth]{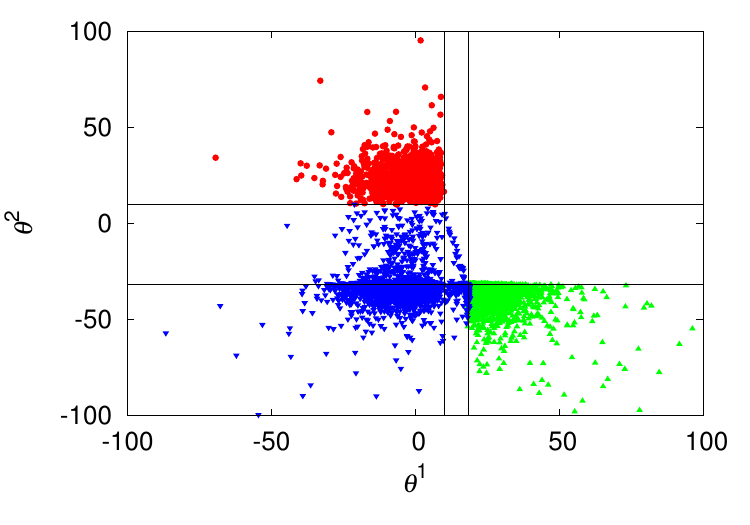}  
\caption{Scatter plot of the thresholds of the agents in the winning group for $\beta = 1/4$. The red circles are the specialists in task 1, the green triangles are the specialists in task 2, and the blue inverted triangles are the generalists.   The horizontal ($\theta^2 = 10$ and $\theta^2 = -31.66$) and vertical ($\theta^1 = 10$ and $\theta^1 = 18.33$) lines are the bounds  (\ref{t_a}) and (\ref{t_b}) for perfect division of labor. The parameters are $M=100$,  $N=100 $, $T=100$, $\delta =1/3$ and  $\sigma^2=1/9$.
 }  
\label{fig:11}  
\end{figure}
%-----------------------------------------------------

An instructive way to summarize and visualize the division of labor  created by the winner-take-all algorithm is through a scatter plot. Accordingly, in Fig.\ \ref{fig:11} we present the scatter plot for $\beta = 1/4$. Each symbol in the figure represents the thresholds $\theta_i^1$ and $\theta_i^2$  of each agent in the winning group for $500$ independent runs, so there are $500N$ symbols in total. The results show that the theoretical bounds  (\ref{t_a}) and (\ref{t_b}) for perfect division of labor in the noiseless case  play a key role in  singling out the specialists as well as  distinguishing between the different types of generalists.  For example, agents with $\theta^2 < -31.66$ and $\theta^1 < 10$ are stimulated by both tasks throughout the observation period. Agents with  $\theta^2 < -31.66$ and $10 < \theta^1 < 18.33$ are always stimulated by task 2, but are stimulated by task 1 only after a certain number of steps in  the stimulus dynamics. 
Agents with  $-31.66 < \theta^2 < 10$ and $\theta^1 < 10$ are always stimulated by task 1, but are  stimulated by task 2 only at the beginning   of the stimulus dynamics. Finally, agents with  $-31.66 < \theta^2 < 10$ and $10 < \theta^1 < 18.33$ are stimulated by task 1 only after a certain number of steps of the stimulus dynamics and by  task 2 only at the beginning  of the stimulus dynamics.  Note that a point in an empty region, e.g.  $\theta^2 > 10$ and  $10 < \theta^1 < 18.33$,  would require  the agent to be idle for some time steps before attending to task 1.  However, as  noted above,  maximizing  $w(T)$ requires that all agents are active throughout observation  period.

%-----------------------------------------------------
\begin{figure}[t] 
\centering
 \includegraphics[width=.8\columnwidth]{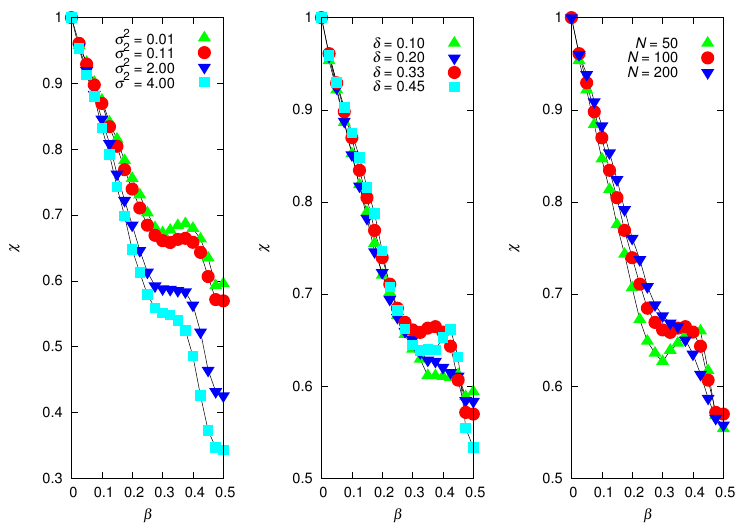}  
\caption{Proportion of specialists  in the winning  group as a function of  the relative importance  $\beta$  of tasks 1 and  2 for (left panel)
$N=100$, $\delta = 1/3$, $\sigma^2 = 0.01, 1/9, 2, 4$,  (middle panel) $N=100$, $\sigma^2 =1/9$, $\delta = 0.1, 0.2, 1/3, 4.5$ and   (right panel) $\delta = 1/3$, $\sigma^2 = 1/9$,  $ N=50, 100, 200$. The symbols are the averages over 500 runs of the winner-take-all algorithm and the lines are guides to eye.  The other parameters are $M=100$ and  $T=100$.
 }  
\label{fig:12}  
\end{figure}
%-----------------------------------------------------

For the sake of completeness, Fig.\ \ref{fig:12} shows the influence of the noise variance $\sigma^2$,  the stimulus increment $\delta$ and the number of agents $N$ on the proportion of specialists $\chi = \nu_1 + \nu_2$ in the winning group. As expected, increasing $\sigma^2$ decreases $\chi$,  since an agent can switch tasks due to noise alone  and thus lose its   specialist status.   This effect is 
more pronounced when the imbalance between  the tasks is small (i.e., $\beta \approx 1/2$).   In this case, the random selection of tasks due to noise does not significantly affect   the fitness of the group. In fact,  for $\beta=1/2$  (and large $T$ and $N$), this strategy maximizes the fitness $w(T)$. Furthermore,   Fig.\ \ref{fig:12} shows that changes in $\delta$ have little quantitative effect  on the proportion of specialists, although they produce somewhat puzzling results  for intermediate values of $\beta$. In principle, we would expect that increasing $\delta$ would favor the generalists, since  large stimulus values increase  the likelihood that the agents will  be stimulated by both  tasks, but this expectation is  only fulfilled for $\beta \approx 1/2$. For small $\beta$,  the group selection pressure to optimize $w(T)$  successfully counteracts the growth of the generalists, and for intermediate $\beta$, the trade-off between these influences  leads to a complicated non-monotonic dependence of $\chi$ on $\delta$ and $\beta$.  Changes in the number of agents $N$ lead to a similarly  complicated trade-off with the group selection pressure,  with small quantitative effects on the proportion of specialists, especially for large $N$. 

We choose $\beta$ as the leading independent  variable in our analysis of the emergence of division of labor in the structured population scenario because it is an  exogenous parameter that appears in the fitness function (\ref{w}) and enforces, through group selection,  a  predetermined distribution  of work across tasks. In fact, since $\beta$ is the only parameter in $w(T)$, we can consider it as a measure of the influence of group selection.  

\section{Discussion}\label{sec:disc}

The main thrust of our paper is that the group fitness $w(T)$, given by equation (\ref{w}),  which imposes the  external demands on the distribution of work across tasks  while seeking to maximize the total work performed by the group, leads  to an (imperfect) division of labor.  This result  is very different from a previous study that found no division of labor at all \cite{Duarte(2012)}. In fact, the appealing feature of  $w(T)$ is that it does not tell us how tasks should be distributed among the agents, which is precisely the question of  division of labor, but rather the total work performed on each task. This raises the question of the degeneracy (or quasi-degeneracy) of the global fitness optimum $w_{opt}$, given by equation (\ref{wopt}). 

%-----------------------------------------------------
\begin{figure}[t] 
\centering
 \includegraphics[width=.8\columnwidth]{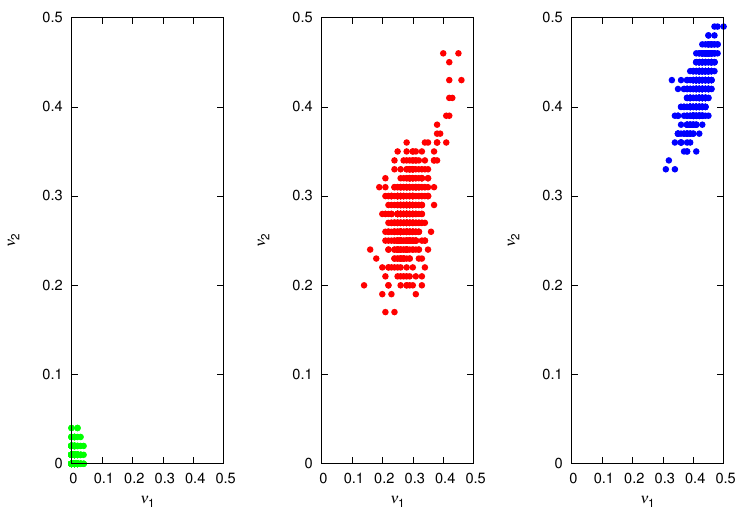}  
\caption{Scatter plots of the proportion of agents in the winning group that work exclusively on task 1 ($\nu_1$) and  that work exclusively on task 2 ($\nu_2$) for $\beta = 1/2$ and initial thresholds  (left panel) $\theta_i^a= 0~\forall i,a$, (middle panel)  $\theta_i^a= 10~\forall i,a$, and (right panel) $\theta_i^a= 20~\forall i,a$. The initial stimulus values are   $S_a(0)=10~\forall a$.  The  parameters are $M=100$, $N=100$, $T=100$, $\delta = 1/3$ and $\sigma^2 = 1/9$.
 }  
\label{fig:13}  
\end{figure}
%-----------------------------------------------------

For instance,  the perfect division of labor resulting from the prescription (\ref{t_a})-(\ref{t_d})  for the thresholds is  a global optimum of $w(T)$ for all $\beta$.  In addition, 
a group consisting of agents that are stimulated by both tasks for the whole observation period $T$, i.e., $\theta_i^a < S_a(0) + T (\delta - 1/2)$ for $i=1,\ldots,N$ and $a=1,2$, also yields a global optimum for $\beta =1/2$. However, the winner-take-all algorithm finds a distinct global optimum, as summarized in Fig.\ \ref{fig:10}. Since the results in this figure   only  show the average proportions of specialists on tasks 1 and 2, we must rule out the possibility that individual runs of the group selection dynamics lead to all-specialist (perfect division of labor) or all-generalist (no division of labor) solutions. Accordingly, Fig.\ \ref{fig:13} shows the proportion of agents in the winning group that work exclusively on task 1 and  that work exclusively on task 2  for $\beta=1/2$ and  500 independent runs of the winner-take-all algorithm. The result shown in the middle panel confirms that the averages shown  in Fig.\ \ref{fig:10} are indeed representative  of the individual runs. Furthermore, the scatter plot allows us to see the dispersion of the results around the mean $\nu_1=\nu_2 \approx 0.285$.

So what biases the winner-take-all algorithm toward the imperfect division of labor solution? We recall that to ensure the key assumption  that the  metapopulation is homogeneous at the beginning of the group dynamics we set $S_a(0)= \theta_i^a=10$  for all agents $i$ and tasks $a$. We have verified that, not surprisingly, the particular numerical value assigned  to the stimulus and thresholds  is inconsequential. The left and right panels of Fig.\ \ref{fig:13} answer our question: if we set $\theta_i^a \ll S_a(0)$,  the algorithm converges quickly   to the all-generalist solution, whereas if we set  $\theta_i^a \gg S_a(0)$, it gets somewhat closer to the all-specialist solution.  This is a nice result, showing that the  winner-take-all algorithm in itself  does not introduce a bias towards certain optima of the fitness $w(T)$: the bias is in the initial setup of the homogeneous groups.  However, we emphasize that this strong degeneracy of the fitness optimum $w_{opt}$ occurs only for $\beta=1/2$. For other values of $\beta$, the choice of the initial uniform thresholds does not significantly affect   the division of labor, since in this case the all-generalist solution is not optimal.  In this sense, the neutral choice $ \theta_i^a = S_a(0)$ used throughout our study does not affect our conclusions.  In particular, it seems impossible to reach the all-specialist solution starting from a homogeneous group composition, since the algorithm is trapped in equally optimal solutions  with imperfect division of labor. Nevertheless, the proportion of  specialists in these solutions is quite significant  (see Fig.\ \ref{fig:12}), so we can claim that group selection  with fitness $w(T)$ can explain the emergence of specialists.

Finally, we note that since the nature of the optimization problem is such that the the fitness increments in each group generation are very small, a nongenetic strategy \cite{Baluja(1995),Maass(2000)}, in which the absolute rather than the relative value of the fitness determines the success of the groups, is necessary to reach  the global fitness  optimum $w_{opt}$. However, this is not really  an issue here since  the reproduction of groups, which are used as proxies for communities in our study, is more likely to follow the wild rules of economic  than of biological competition.

\bigskip

\acknowledgments
JFF was  supported in part 
 by Grant No.\  2020/03041-3, Fun\-da\-\c{c}\~ao de Amparo \`a Pesquisa do Estado de S\~ao Paulo 
(FAPESP) and  by Grant No.\ 305620/2021-5, Conselho Nacional de Desenvolvimento 
Cient\'{\i}\-fi\-co e Tecnol\'ogico (CNPq). PRAC was partially supported by Conselho Nacional de Desenvolvimento Cient\'{\i}fico e Tecnol\'ogico (CNPq) under Grant No. 301795/2022-3, and acknowledge
financial support from Coordena\c{c}\~ao de Aperfei\c{c}oamento de
Pessoal de N\'{\i}vel Superior (CAPES) (Project No.\ 0041/2022). VMO acknowledges financial support from Conselho Nacional de Desenvolvimento Cient\'{\i}fico e Tecnol\'ogico (CNPq) under Project No.\ 404057/2021-7.

\bibliographystyle{Frontiers-Harvard}

\end{document}